\documentclass[journal]{IEEEtran}
\usepackage[left=0.625in,right=0.625in,top=0.75in,bottom=1in]{geometry}
\usepackage{ifpdf}
\usepackage{cite}
\usepackage{algorithmic}
\usepackage[ruled, lined, linesnumbered, commentsnumbered, longend]{algorithm2e}
\usepackage{array}
\usepackage{dblfloatfix}
\usepackage{url}
\usepackage{amsmath}
\usepackage{ragged2e}
\usepackage{graphicx}
\usepackage{verbatim}
\usepackage{caption}
\usepackage{float}
\usepackage{bbm}
\usepackage{subcaption}
\usepackage{titlesec}
\usepackage{bbm}
\usepackage{xcolor}

\begin{document}
\title{\fontsize{20}{30}\selectfont Age of Information Minimization using Multi-agent UAVs based on AI-Enhanced Mean Field Resource Allocation}
\author{Yousef~Emami,~\IEEEmembership{Member,~IEEE,}
        Hao~Gao,~\IEEEmembership{Student Member,~IEEE,}
        Kai~Li,~\IEEEmembership{Senior Member,~IEEE,}
        Luis~Almeida,~\IEEEmembership{Senior Member,~IEEE,}
        Eduardo~Tovar,~\IEEEmembership{Member,~IEEE,}
        and~Zhu~Han,~\IEEEmembership{Fellow,~IEEE}
\thanks{Copyright (c) 2024 IEEE. Personal use of this material is permitted. However, permission to use this material for any other purposes must be obtained from the IEEE by sending a request to pubs-permissions@ieee.org.} 
}

\maketitle
\begin{abstract}
Unmanned Aerial Vehicle (UAV) swarms play an effective role in timely data collection from ground sensors in remote and hostile areas. Optimizing the collective behavior of swarms can improve data collection performance. This paper puts forth  a new mean field flight resource allocation optimization to minimize age of information (AoI) of sensory data, where balancing the trade-off between the UAVs’ movements and AoI is formulated as a mean field game (MFG). The MFG  optimization yields an expansive solution space encompassing continuous state and action, resulting in significant computational complexity. To address practical situations, we propose, a new mean field hybrid proximal policy optimization (MF-HPPO) scheme to minimize the average AoI by optimizing the UAV's trajectories and data collection scheduling of the ground sensors given  mixed continuous and discrete actions. Furthermore, a long short term memory (LSTM) is leveraged in MF-HPPO to predict the time-varying network state and stabilize the training. Numerical results demonstrate that the proposed MF-HPPO reduces the average AoI by up to 45\% and 57\% in the considered simulation setting, as compared to multi-agent deep Q-learning (MADQN) method and non-learning random algorithm, respectively.
\end{abstract}

\IEEEpeerreviewmaketitle
\section{Introduction} \label{sec:1}
Thanks to high mobility, unmanned aerial vehicles (UAVs) are widely used in search and rescue\cite{10.1145/3571728}, monitoring and surveillance\cite{li2019energy},\cite{li2020joint}, aerial data relay\cite{datarelay},\cite{8854903}, construction\cite{construction} and parcel delivery\cite{cargo}. A large number of ground sensors can be deployed in a target area to monitor the condition. Multiple UAVs can be employed to gather sensory data\cite{li2022data}. UAVs-assisted data collection offers several advantages for the data collection in remote and human-unfriendly environments. Specifically, UAVs have the ability to reach areas that are difficult to access by humans, making data collection more efficient and cost-effective. This reduces safety risks since the use of UAVs eliminates the need for human intervention in hazardous environments. Due to mobility, UAVs are able to cover large areas, which reduces the time and resources required for data collection \cite{li2021continuous},\cite{li2022deep}.
\par
In UAVs-assisted sensor networks, Age of Information (AoI) is typically used to describe the freshness of the sensory data \cite{kaul2012real}, which refers to the elapsed time between the generation of a piece of sensory data at a ground sensor and the receipt of that data at the UAV. Therefore, AoI takes into account not only the time it takes for data to be transmitted, but also any delays that may occur in the network. When the UAV flight is not properly controlled, the UAV can move away from the ground sensor, causing the AoI of the sensor to be extended and resulting in data expiration. 
In addition, AoI of the ground sensors' data can be different from each other, since the sensory data generation is impacted by natural conditions monitored \cite{aggregator}. The instantaneous knowledge of data generation rate and channel conditions of the ground sensors are not available or can only be partially observed by the UAV in practical systems. Therefore, jointly optimizing the cruise control of the UAVs and communication schedules of the ground sensors to minimize AoI is non-trivial. Moreover, swift movements of the UAV results in poor channel conditions and fast signal attenuation, giving rise to frequent data retransmissions and a prolonged AoI. In contrast, slow movements of the UAV extend flight time, thereby extending AoI of the ground sensors. Meanwhile, the UAV should consider the movement of other UAVs to minimize the average AoI by coordinating their velocities. Therefore, finding an equilibrium solution among UAVs to optimize the velocity and reduce the average AoI is essential. 

The decentralized approach is particularly relevant in
situations where the UAVs have limited information about the actions of others, in terms of the trajectory, flight speed, and the scheduled ground sensors. Game theory can be leveraged to design decentralized control and determine the equilibrium in UAV networks \cite{8723552}. However, traditional game theory approaches that focus on interactions between a finite number of UAVs (or players) can become intractable when dealing with a large number of UAVs, as the complexity of solving such games grows exponentially with the number of UAVs. Diverging from traditional game theory, mean field game (MFG) offers a scalable framework to addressing the proposed joint optimization of the cruise control and communication schedules, by approximating the interactive behavior of a large number of UAVs using a continuum or mean field, thereby considerably reducing computational complexity. Moreover, MFG allows the UAVs to make decisions based on the behavior of the overall swarms rather than on the actions of individual UAVs. 

In this paper, we propose a cruise control optimization of the UAV swarms based on MFG to minimize AoI, while balancing the trade-off between the UAVs' movements and AoI. In our MFG, the optimal velocities of the UAVs are determined by a Fokker-Planck-Kolmogorov (FPK), which describes an evolution of the mean field for achieving an equilibrium of the optimal velocities of the UAVs. However, the proposed MFG is difficult to be solved online in practical scenarios, since the instantaneous knowledge of the UAV's cruise control decision and AoI is hardly known by other UAVs. Furthermore, the MFG for the flight resource allocation optimization is formulated as a multi-agent Markov Decision Process (MMDP), where  network states consist of AoI of the ground sensors and waypoints of the UAV swarm. The action space in the MMDP contains the waypoints and velocities in a continuous space and transmission schedule in a discrete space. Based on the formulated MMDP, we propose a mean field hybrid proximal policy optimization (MF-HPPO) to minimize AoI of the ground sensors, where the UAV swarms learn each other's decisions of cruise control and transmission schedules of the ground sensors.  

In summary, UAV swarms are employed to collect time critical sensory data. Time-critical data collection is influenced by the velocity of the UAVs and their coordinated interactions in the swarms, which can be modeled using MFG. This raises the importance of an age-optimal cruise control based on MFG for UAVs.  However, determining the equilibrium online is difficult in practical scenarios, and thus we propose MF-HPPO that highly reduces the complexity while minimizing the average AoI. The contributions of this paper are listed as follows:
\begin{itemize}
    \item We novelly formulate the MFG optimization with a large number of UAVs to address the trade-off between the cruise control of the UAVs and AoI. Due to the high computational complexity of the MFG, MF-HPPO is proposed  to minimize the average AoI, where the state dynamics are learned and the actions of the UAVs are optimized in a mixed discrete and continuous action space.
    \item To capture temporal dependencies of the cruise control while improving the learning convergence, a new long short term memory (LSTM) layer is developed with the proposed MF-HPPO to predict the time-varying network states, i.e., AoI and UAVs' waypoints.
    \item Numerical results demonstrate that MF-HPPO achieves fast convergence (less than 200 iterations). Meanwhile, our proposed MF-HPPO  witnesses 45\% and 57\% reduction in AoI as compared to a multi-agent deep Q-learning (MADQN) method (which performs  trajectory planning in the discrete space) and an existing non-learning random algorithm, respectively.

\end{itemize}
\par
The rest of this paper is organized as follows:
In Section \ref{sec:2}, we present the state-of-the-art in intelligent flight resource allocation, with a focus on MFG and time-critically. In Section \ref{sec:3}, we formulate the channel model as well as the AoI in the UAVs-assisted sensor network. In Section \ref{sec:4}, we formulate the flight resource allocation of the UAV swarm as the MFG to minimize the AoI. Section \ref{sec:5} develops the proposed MF-HPPO, to jointly optimize the cruise control of multiple UAVs and data collection scheduling. Section \ref{sec:6} presents the implementation of the proposed MF-HPPO in Pytorch as well as performance evaluation. Finally, Section \ref{conc} concludes this paper.
\section{Related Work} \label{sec:2}
\subsection{Mean field flight resource allocation} \label{sub:1}
In \cite{chen2020mean}, the authors explore energy-efficient control strategies for UAVs that provide fair communication coverage for ground users. The UAV control problem is modeled as an MFG and a mean-field trusted region policy optimization (TRPO) algorithm is studied to design the UAVs' trajectories. In \cite{li2020downlink}, the authors apply the MFG theory to the downlink power control problem in ultra-dense UAV networks to improve the network's energy efficiency. Due to the complexity of the MFG, a DRL-MFG algorithm is developed to learn the optimal power control strategy. \cite{service} studies the task allocation in cooperative mobile edge computing and a mean field guided Q-function is formulated to reduce the network latency. MFG and deep reinforcement learning (DRL) are integrated to guide the learning process of DRL according to the equilibrium of MFG. In \cite{sun2020joint}, the authors model the trajectory planning and power control for heterogeneous UAVs as an MFG, aiming to reduce energy consumption. A mean field Q-learning is studied to find the optimal solution. In \cite{uavpositionm}, the authors study UAV-assisted ultra-dense networks, where each UAV can adjust its location to reduce the AoI. They formulate the problem as an MFG and apply a deep deterministic policy gradient (DDPG)-MFG algorithm to find the mean field equilibrium. In \cite{li2020downlink}, downlink power control for a large number of UAVs is suggested to enhance the energy efficiency  by learning the optimal power control policy. MFG is used to model the power control problem of the UAV network, where each UAV tries to enhance the energy efficiency by adjusting its transmit power. Then, due to the complexity of solving the formulated MFG, an effective DRL-MFG algorithm is suggested to learn the optimal power control strategy. 

Although, DRL-based solutions mainly used, the following works adopt numerical solutions. In \cite{xue2018adaptive}, the focus is on adaptive coverage problem in emergency communication system, where multiple UAV act as aerial base stations to serve randomly distributed users. The problem is formulated using discrete MFG, each UAV aim to reduce its flight energy consumption and increase the number of users it can serve. Finally, optimal control and state of each UAV is computed. In \cite{xu2018discrete}, a discrete MFG is formulated to address joint adjustment of power and velocity for a large number of UAVs that act as aerial base stations. Decentralized  control laws are developed, and mean field equilibrium is analyzed. In \cite{velocitycontrol}, the authors present an energy-efficient velocity control algorithm for a large number of UAVs based on the MFG theory. The velocity control of the UAVs is modeled using a differential game in which energy and delay are balanced by using an original double mixed gradient method. In \cite{liu2022joint}, a multi-UAV enabled mobile Internet of Vehicles (IoV) model is designed, to enable the UAVs to track the movements of the vehicles. A joint optimization problem is formulated to improve the total system throughput over the flight time by jointly adjusting vehicle communication scheduling, UAV power allocation, and UAV trajectory. In \cite{liu2022fair}, a multi-UAV-enabled IoT is investigated to improve the minimum energy efficiency of each UAV by jointly adjusting communication scheduling, power allocations, and trajectories of the UAVs.

\subsection{Time-critical flight resource allocation} \label{sub:2}
In \cite{9750860}, the authors consider ground sensors with limited energy and apply airborne base stations to collect sensory data. Each UAV's task is decomposed into energy transfer and fresh data collection. A centralized multi-agent DRL based on DDPG is developed to adjust the UAV trajectories in a continuous action space, to reduce the AoI of the ground sensors. In \cite{9909005}, the authors study UAV-assisted sensor networks where multiple UAVs cooperatively conduct the data collection to reduce the AoI. The trajectory planning is formulated as a decentralized partially observable Markov Decision Process (Dec-POMDP). A multi-agent DRL is studied to find the optimal strategy. In \cite{hu2019distributed} and \cite{hu2020cooperative}, the authors develop the trajectory planning for multiple UAVs that perform cooperative sensing and transmission, aiming to reduce the AoI. In \cite{9285215}, ground sensors sample and upload data in a UAV-assisted IoT network. PPO is used to explore the optimal scheduling policy and altitude control for the UAV to reduce the AoI. In \cite{sun2021aoi}, a data collection scheme characterized by AoI and energy consumption in a UAV-assisted IoT network is investigated. The average AoI, and energy consumption of propulsion and communication are reduced by adjusting the UAV flight speed, hovering waypoints, and bandwidth allocation for data collection using a TD3-based approach. In \cite{liu2022fair} a multi-UAV enabled IoT is investigated to maximize the minimum energy efficiency of each UAV by jointly optimizing communication scheduling, power allocations, and trajectories of the UAVs. In \cite{10257579}, a UAV-assisted IoT network is investigated in the presence of eavesdroppers. The communication UAV and the jamming UAV cooperate to collect data from IoT devices. A TD3-based solution is developed to reduce the AoI by adjusting the UAV trajectory, computations, and spectrum resource allocation strategies.

Although, DDPG and PPO used to adjust continuous and discrete actions to reduce AoI, the following works use DQN and QMIX to adjust discrete actions. In \cite{9815722}, the authors investigate UAV-assisted IoT networks where multiple UAVs relay data between sensors and base station. A DQN-based trajectory planning algorithm is presented to reduce the AoI. In \cite{abd2019deep}, ground sensors with limited energy are used to observe various physical processes in the context of a UAV-assisted wireless network. The trajectory and scheduling policy are adjusted to reduce the weighted sum of AoI, and a DQN-based solution is applied to obtain the best strategy.  In \cite{zhou2019deep}, trajectory planning of the UAV is performed to reduce the AoI in a UAV-assisted IoT network. The problem is formulated as an MDP, and a DQN-based algorithm is studied to find the optimal trajectories of the UAV. In \cite{tong2020deep}, a UAV-assisted data collection for ground sensors is studied, where the UAV with limited energy is dispatched to collect sensory data. The UAV's trajectory is adjusted to reduce the average AoI and keep the packet loss rate low. The trajectory planning is formulated as an MDP while DQN is applied to design the UAV's trajectory. In \cite{liu2021average}, a UAV-assisted wireless network with an energy supply is used, where the UAV performs wireless energy transmission to ground sensors, and the sensors transmit data to the UAV using the harvested energy. A DQN-based trajectory planning algorithm is presented to reduce the average AoI by adjusting the trajectory, transmission schedule, and harvested energy. In \cite{9950310}, a massive deployment of up to a hundred IoT devices is investigated, where multiple UAVs serve as mobile relay nodes to reduce the AoI and energy consumption. A DQN-based solution is developed to jointly reduce the AoI and energy consumption of the devices by adjusting the trajectory and scheduling policy. In \cite{10188891}, a UAV-assisted sensor network is investigated, where the UAV flies between the resource-constrained sensors and collects their status updates. A DQN-based solution is developed to jointly adjust the trajectory of the UAV and the scheduling of the sensors to reduce the Age of Synchronization (AoS), which considers both the freshness and the content of the information. In \cite{10293964}, a massive IoT communication scenario is investigated where a UAV swarm collects fresh information from IoT devices and provides better coverage and LoS for the IoT network. To mitigate high-dimensional problems with high complexity, a multi-agent DRL based on DQN is developed. In \cite{10065524}, the problem of optimal data collection in IoT networks with multiple cooperative UAVs is investigated. Kinematic, energy, trajectory, and collision avoidance constraints are considered. To achieve this goal, a QMIX-based algorithm is developed to reduce the total average AoI by jointly adjusting the trajectories of the UAVs and the scheduling of the sensors. 

Most of the works, in Section \ref{sub:1}, formulate an MFG for multiple UAVs to address energy efficiency. Authors in \cite{uavpositionm} formulate an MFG to minimize AoI and suggest DDPG-MFG in continuous action space to find the  optimal solution.
\begin{figure} [H]
        \centering
        \includegraphics[ width=3in, height =2in]{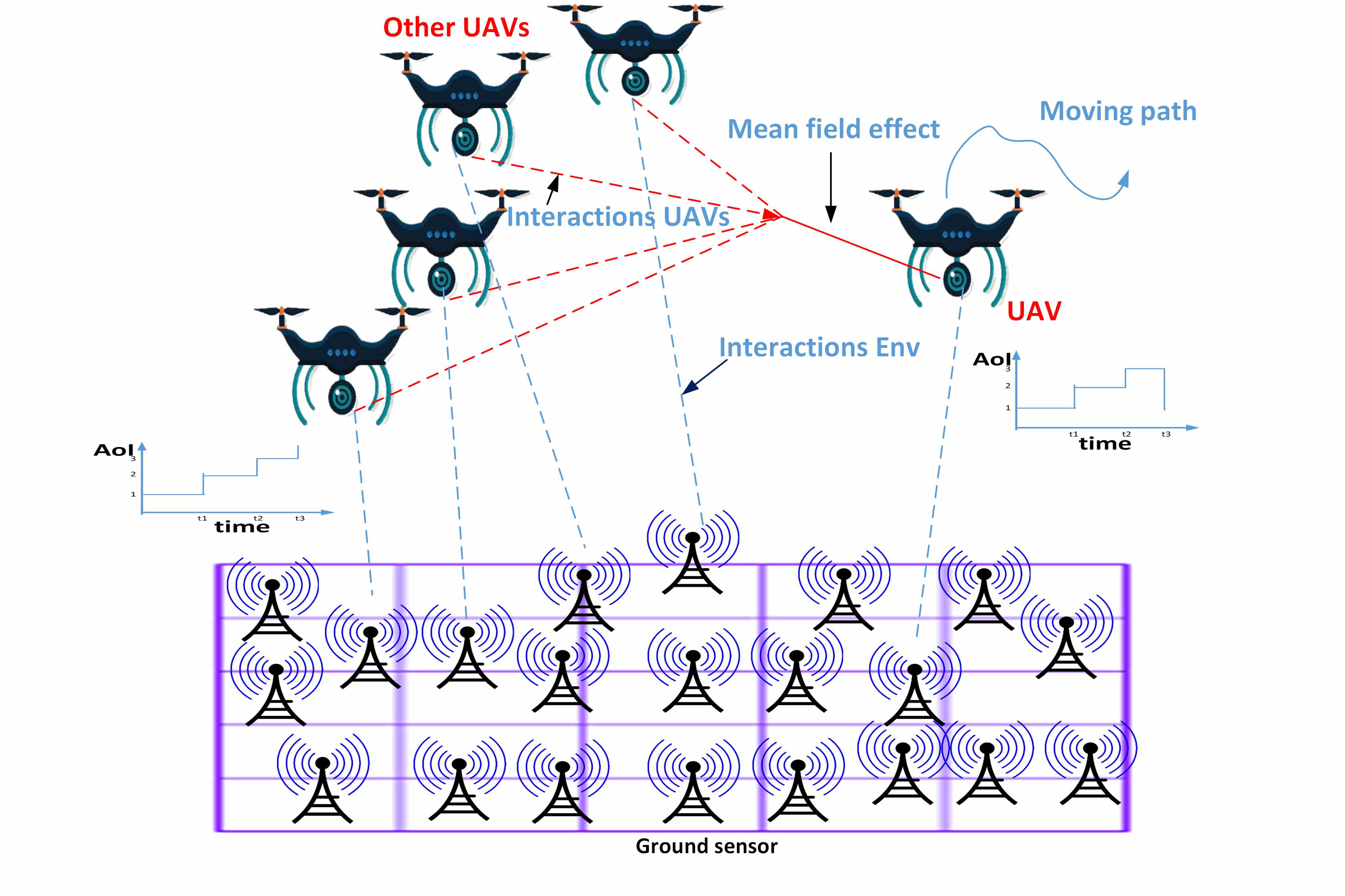}
    	\caption{Mean field representation of UAVs-assisted sensor networks.}
    	\label{fig:forest}
\end{figure}
The works in Section \ref{sub:2} investigate resource allocation to reduce AoI, however, actions are adjusted in continuous or discrete action spaces. In contrast, in this paper, we novelly formulate the MFG optimization with a large number of UAVs to address the trade-off between the cruise control of the UAVs and AoI. Due to the high computational complexity of the MFG, MF-HPPO is proposed to minimize the average AoI, where the state dynamics are learned and the actions of the UAVs are optimized in a mixed discrete and continuous action space.
\section{System Model}  \label{sec:3}
In this section, we present the system model of the considered UAVs-assisted sensor network. Notations used in this paper are summarized in Table \ref{table:1}. 
The system consists of \emph{$I$} UAVs, $\emph{$i$}\in[1,\emph{$I$}]$ and \emph{$J$} ground sensors, $\emph{$j$}\in[1,\emph{$J$}]$ in which the ground sensors are deployed in a target region. The UAVs are employed to patrol in the target zone while collecting the sensory data. Fig. \ref{fig:forest} depicts an example of UAVs-assisted sensor network along with mean field representation. With the increase in the number of UAVs in Fig. \ref{fig:forest} the interactions between them become complex and can dominate the overall behavior of the system. MFG designed to deal with the optimal control problem involving a large number of players. It has unique characteristics suitable for UAV swarm and modelling these interactions. Each UAV seeks to minimize the AoI according to the actions of other agents surrounded. As depicted, the UAV consider the mean field effect of the other UAVs, which represents the collective behavior of the UAVs in the system. The coordinates $(x_i,y_i,z_i)$ and $(x_j,y_j,0)$ represent the position of  UAV \emph{$i$} and ground sensor \emph{$j$}, respectively. The UAVs fly to the ground sensors, collect sensory data, and then their operation is terminated. The UAVs fly at a constant altitude, represented by $\zeta_i(t) = (x_i,y_i,z)$. The distance between ground sensor \emph{$j$} and UAV \emph{$i$} is $\sqrt{(x_i-x_j)^2+(y_i-y_j)^2+z^2}$. For the safety of the UAV during flight by preventing it from exceeding the maximum safe speed or stalling, we denote the maximum and minimum velocity of the UAV as $v_{max}$ and $v_{min}$, respectively. The lower bound of the speed, i.e., $v_{min}$, is set to 0.
\begin{table}[h!]
\centering
\caption{Notation and Definition}
\begin{tabular}{|c|c|} 
 \hline
 \textbf{Notation} & \textbf{Definition}\\  
 \hline
 \emph{$J$} & number of ground sensors  \\ 
 \hline
 \emph{$I$} & number of UAVs \\
 \hline
  $h_j^i(t)$ & channel gain between device j and UAV i \\
 \hline
 $\zeta_i(t)$ & location of the UAV on its trajectory\\
 \hline 
 $v_i(t)$ & velocity of UAV i\\
 \hline
 $v_{max},v_{min} $ & the maximum and minimum velocity of UAV i\\
 \hline
 $M$ & number of episodes \\
 \hline
 $L$ & length of each episode \\
 \hline
 $\gamma$ & discount factor \\
 \hline
 $\eta$ &  learning rate \\ 
 \hline
 $D$ & buffer size \\
 \hline
 $B$ & mini-batch size \\
 \hline
 $a_i$ & action of UAV $i$ \\
 \hline
 $o_i$ & mean field of UAV $i$ \\
 \hline
 $a_i^c$ & continuous action of UAV $i$ \\
 \hline
 $a_i^d$ & discrete action of UAV $i$\\
 \hline
 $s_{\alpha,i}$ & state of UAV $i$\\
 \hline
 $E[..]$ & mathematical expectation\\
 \hline
 $A$  & advantage function \\
 \hline
 $\theta$ & network parameter \\
 \hline
 $\pi$ & policy \\
 \hline
 $\pi^c$ & continuous policy \\
 \hline
 $\pi^d$ & discrete policy \\
 \hline
 $\sigma$ & diffusion coefficient  \\
 \hline
 $W$ & weiner process \\
 \hline
 $H$ & entropy \\
 \hline
 \end{tabular}
\label{table:1}
\end{table}
We consider that UAV \emph{$i$} moves in low attitude for data collection, where the probability of LoS communication between UAV \emph{$i$} and ground sensor \emph{$j$} is given by \cite{lap} 
\begin{equation} \label{eq:2}
    \Pr{_{LoS}}(\varphi_j^i)=\frac{1}{1+a \exp(-b[\varphi_j^i-a])}
\end{equation}
where \emph{$a$} and \emph{$b$} are constants, and  $\varphi_j^i$ denotes the elevation angle between  UAV \emph{$i$} and ground sensor \emph{$j$}. The elevation angle of the path loss model is given by
\begin{equation}
\varphi = \tan^{-1}\left(\frac{h}{d}\right),
\end{equation}
where $h$ is the height of the UAV above the ground and $d$ is the horizontal distance between the UAV and the ground sensor. By changing $d$, the horizontal distance between the UAV and the ground sensor, we can influence the path loss of our UAV communication through the elevation angle.
Moreover, path loss of the channel between  UAV \emph{$i$} and device \emph{$j$} can be modeled by \cite{9531342}
\begin{multline} \label{eq:3}
  \gamma_j^i=\Pr{_{LoS}}(\varphi_j^i)(\eta_{LoS}-\eta_{NLoS})+20 \log \left(r \sec(\varphi_j^i)\right)+ \\20 \log(\lambda)+20 \log \left(\frac{4\pi}{v_c}\right)+\eta_{NLoS}    
\end{multline}
where \emph{$r$} is the radius of the radio coverage of UAV \emph{$i$}, \emph{$\lambda$} is the carrier frequency, and \emph{$v_c$} is the speed of light. $\eta_{LoS}$ and $\eta_{NLoS}$ are the excessive path losses of LoS or non-LoS, respectively.

To characterize the freshness of the collected sensory data at the UAV, AoI is defined as the time that has passed since ground sensor generates the latest information. The AoI of ground sensor $j$ that generated a data packet at $t_j$ and collected by UAV $i$ at $t_i$ is given by
\begin{equation} \label{eq:333}
AoI^i_j(t)=t_i-t_j.
\end{equation}
\noindent According to (\ref{eq:333}), it can be also known that maintaining a low $AoI^i_j(t)$ is critical for improving the effectiveness and timeliness of the sensory data, reducing the response time, and providing real-time information for decision-making at the UAVs. n UAVs-assisted sensor networks, the AoI can be effected by  the path loss condition: Decreased path loss attenuates signal deterioration during the conveyance from ground sensors to UAVs, culminating in enhanced signal strength, expedited data transfer, diminished error rates, and subsequently, a reduced necessity for retransmissions. Such efficacious communication markedly diminishes the AoI and guarantees the contemporaneity and pertinence of the data for real-time applications. Conversely, elevated path loss results in attenuated signals, escalated error frequencies, increased retransmissions, and ultimately, an augmented AoI, thereby impacting the promptness and dependability of the data.
\section{Problem Formulation} \label{sec:4}
In this section, we formulate the MFG optimization with a large number of UAVs to address the trade-off between the cruise control of the UAVs and AoI. We also explore the FPK equation to determine the optimal velocities of the UAVs while characterizing the collective behavior of the UAVs. We begin with optimal control formulation in Section \ref{sub:41} and then proceed with MFG formulation in Section \ref{sub:42}.
\subsection{Optimal Control Formulation} \label{sub:41}
We derive the state dynamics and cost function, then we formulate the velocity control problem using the optimal control theory.
\subsubsection{Time-varying Dynamics of Network States}
Let $\zeta_i(t)$ denote the position of the UAV $i$ at time $t$ and $v_i(t)$ denotes the velocity. 
According to Newton's laws of motion \cite{waldrip2013explaining}, the location dynamics of UAV $i$ can be expressed by 
\begin{equation} \label{eq:4}
    d\zeta_i(t)=v_i(t) dt+ \sigma dW_i(t)
\end{equation}
where $W_i(t)$ is a standard Wiener process \cite{morters2010brownian} with a diffusion coefficient $\sigma$. 
\subsubsection{Cost Function}
Each UAV intends to optimize its velocity to minimize the cost function. Our cost is defined as the average AoI of all ground sensors.
The average AoI can be computed as:
\begin{equation} \label{eq:7}
    c(t)=\frac{1}{IJ}\Sigma_{j=1}^J \Sigma_{i=1}^I AoI^i_j(t). 
\end{equation}
\subsubsection{Velocity Control Problem Formulation} \label{sec:optimal}
Given a period of time $T$ regarding the data collection, the velocity of UAV $i$ at $t$, denoted as $v_i^*(t)$, is optimally controlled to minimize $c(t)$, which gives:
\begin{equation} \label{eq:8}
v_i^*(t)= \underset{v_i(t)}{\arg \min} E\int_0^T c(t)dt),
\end{equation}
\[ s.t.\ (\ref{eq:4}).\]
Equation (\ref{eq:8}) is the integral of c(t) on the given time range (0,T), while c(t) is the average AoI of all UAVs defined in (\ref{eq:7}). And the definition of AoI is defined in (\ref{eq:333}). In summary, physical meaning of  (\ref{eq:8}) is to find the optimal velocity of UAVs to minimize the accumulated average AoI in the time range (0,T). Constraint (5) is a differential constraint of a velocity control problem. To be more precise, the Newtonian motion function describes the change of locations of UAVs w.r.s.t their velocities. The Brownian motion, $\sigma$,  represents the random effects, which might influence the locations of UAVs. The use of game theory in our model, as opposed to direct optimization, is crucial due to the inherent interdependencies between the decisions of multiple UAVs that are not explicitly captured in (\ref{eq:8}). While equation (\ref{eq:8}) represents a stand-alone minimization problem for the control strategy \( v_i(t) \) of a single UAV, the choice of \( v_i(t) \) by one UAV in a real-world scenario with multiple UAVs implicitly affects the operational efficiency and AoI outcomes of the others. 
To determine $v_i^*(t)$ in (\ref{eq:8}), classical game theories, such as differential game, fails to capture the aggregate behavior of all the UAVs. Differential game assumes each agent's movement is independent of others. This assumption fails to capture the fact that a large number of UAVs' trajectories decisions are influenced by the aggregate behavior of all the UAVs, thus hardly minimizing the average AoI, $c(t)$. 

We novelly extend MFG to capture the impact of the aggregate behavior of the UAVs, in terms of cruise control. The MFG models the aggregate decision of UAVs as a probability distribution, rather than focusing on the actions of individual UAVs. This recognizes that the cruise control of each UAV is influenced by the behavior of all other UAVs. Moreover, the formulated MFG is defined to minimize $c(t)$ given a large number of UAVs, which classical game theory struggles with due to the computational complexity of solving for the equilibrium. 
\subsection{MFG Problem Formulation} \label{sub:42}
We reformulate the optimal cruise control problem in (\ref{eq:8}) into a cooperative MFG problem. The computational complexity of the system is greatly reduced by formulating an MFG, since a large number of interactions with other agents is converted into an interaction with the mass.
The interaction between each UAV with the other UAVs is modeled as a mean-field term, which is denoted by $m(\zeta(t))$. The mean-field term is the distribution over agents´ state space or control to model the overall state and control of them. We can measure the state and control of all agents in an MFG using the mean-field term.\\

Given dynamics, $\zeta_i(t)$, the mean-field term of $m(\zeta(t))$ can be denoted by
\begin{equation} \label{eq:9}
 m(\zeta(t))=\lim_{I\to\infty} \frac{1}{I} \Sigma_{i=1}^{I} \mathbbm{1} \{\zeta_i(t)=\zeta(t)\},   
\end{equation}
where $\mathbbm{1}$ is an indicator function which returns 1 if the given
condition is true, or 0, otherwise.

Given $m(\zeta(t))$, the state dynamics, cost function and FPK equation can be defined as:\\
\begin{itemize}
    \item  \textbf{State dynamics:} The state dynamics of each UAV  can be expressed by

       \begin{equation} \label{eq:10}
            d\zeta(t)=v(t) dt+ \sigma dW(t).
       \end{equation}
    \item  \textbf{Cost function:} The mean-field term affects the running cost function of each UAV. The average AoI of the all UAVs is computed by
\begin{equation} \label{eq:11}
    c(v(t),m(\zeta(t)))=\int c(v(t))\cdot m(\zeta(t)) d\zeta.
\end{equation}
Mathematically, the cost function can be written by
\begin{equation} \label{eq:12}
    J(v(t),m(\zeta(t))))=\int_{t=0}^T c(v(t),m(\zeta(t)) dt.
\end{equation}
If the UAV move quickly, lead to poor channel condition and retransmissions thereby AoI prolongs. In contrast, slow movement of the UAV, may prolong the AoI of the ground sensors because the data are not collected in time. The cost function addresses these trade-offs and find the optimal velocity to balance these objectives.
\item  \textbf{Focker-Planck equation:} Based on (\ref{eq:10}) we develop the FPK equation. The FPK equation governs the evolution of the mean field function of UAVs and given by:
\begin{equation} \label{eq:13}
    \partial_tm(\zeta(t))+\nabla_{\zeta}m(\zeta(t))\cdot v(t)- \frac{\sigma^2}{2}\nabla_{\zeta}^2m(\zeta(t))=0. 
\end{equation}
See \it{Appendix}.
\end{itemize}
After deriving the state dynamics, cost function, and FPK equation, we now proceed to present the MFG. 

To summarize, the cooperative MFG problem is given by
\begin{equation} \label{eq:14}
\min_{v,m} J(v(t),m(\zeta(t)))
\end{equation}
\[ s.t.\ (\ref{eq:13}).\]
\section{Proposed MF-HPPO} \label{sec:5}
In this section, we present background on PPO in Section \ref{sub:51} and then describe the MFG as an MMDP in Section \ref{sub:52} so that the optimal actions of UAVs can be learned by the proposed MF-HPPO. MF-HPPO is presented in Section \ref{sub:53}, which employs onboard PPO to minimize the average AoI  of the ground sensors. The trajectory and instantaneous speed of the UAVs, and the selection of the ground sensors are optimized in a mixed action space. In Section \ref{sub:54}, an LSTM  layer is developed with MF-HPPO  to capture the long-term dependency of data.
\subsection{Background on PPO} \label{sub:51}
In this work, we use policy-based DRLs because of their superior performance compared to state-of-the-art algorithms. A major issue in this category is update instability, which is rooted in the variability of the step-size parameter for policy optimization. Small steps slow the learning process, while large steps degrade policy performance. TRPO \cite{trust} addresses this issue by defining a trusted region for changes and using a complex second-order method. PPO follows the same approach by presenting a first-order method to overcome the high complexity of TRPO. PPO bounds the policy update within a range as an alternative to the hard constraint of TRPO. PPO has two primary variants: PPO-penalty and PPO-clip. PPO-penalty changes the hard constraint of TRPO to a penalty in objective function. PPO-clip does not have a constraint and use clipping techniques to bound the changes of policy \cite{samir2020online}. The PPO-clip objective function can be written as follows:
\begin{equation} \label{eq:33}
          \begin{split}
             & L^{clip}(\theta)=\min\Bigl(\frac{\pi(a|s)}{\pi_{old}(a|s)}A_{\pi_{old}}(s,a),
             \\
             & g(\epsilon,A_{\pi_{old}}(s,a))\Bigr),
          \end{split}
\end{equation}
where

\begin{equation}
g(\epsilon,A) =
\begin{cases}
  (1+\epsilon)A ,    &    \text{$A \geq 0$},\\
  (1-\epsilon)A ,    & \text{$A \textless 0$}.\\
\end{cases}
\end{equation}
where $\epsilon$ is used to control the clip range. PPO uses Actor-Critic framework in the implementation step. The overall objective function is given by:
\begin{equation} \label{eq:34}
    L^{total}(\theta)=L^{clip}(\theta)-K_1 L^{VF}(\theta)+K_2*H
\end{equation}
where $K_1$ and $K_2$ are loss coefficients, $H$ is entropy. $L^{VF}$ is loss for Critic network. The proposed approach, MF-HPPO, leverages PPO-clip as model-free, on-policy and policy gradient DRL algorithm and is capable to optimize continuous and discrete actions. In the following, we formulate our problem using MMDP and then address it using MF-HPPO. 
\subsection{MMDP Formulation} \label{sub:52}
We reformulate the MFG using MMDP framework to enable the application of PPO for optimizing the actions and minimizing average AoI. By adapting the MMDP framework to our problem, we define the relevant state space, action space, transition probabilities, policy and cost function, thus facilitating an effective solution approach based on MF-HPPO. We define our MMDP as follows.
\begin{itemize}
    \item {\em Agents}: the number of agents, i.e., UAVs is denoted by \emph{I}.
    \item {\em State}: A state $s_{\alpha}$ of the MMDP consists of the positions of UAV $i$, the AoI of ground sensors, i.e, $s_{\alpha}$=\{$\zeta_i(t)$,$AoI^i_j(t):i \in [1,I],j \in [1,J] $\}. All states of the MMDP constitute the state space.
    \item {\em Action}: Each UAV $i$ takes an action $a_i$ that schedules a ground sensor for data transmission and determines the flight trajectory and velocity, i.e, $a_i$ =\{$k_j^i$,$v_i(t)$,$\zeta_i(t)$\}
    \item {\em Policy}: Policy $\pi_i$ is the probability of taking each action of agent i.
    \item {\em State Transition}: The current state $s_{\alpha}$ transit to a new state $s_{\beta}$ according to probability $P(s_{\beta}\mid s_{\alpha},a)$, where $a$ indicates a joint action set that includes the actions of all the UAVs.
    \item {\em Cost}: The immediate cost of the UAVs is $\frac{1}{IJ}\Sigma_{j=1}^J \Sigma_{i=1}^I AoI^i_j(s_{\alpha},a)$.    
   
\end{itemize}
\subsection{MF-HPPO} \label{sub:53}
The proposed MF-HPPO operates onboard at the UAVs to determine their trajectories and  sensor selection. The UAV chooses a sensor and moves to it, then sends out a short beacon message with the ID of the chosen sensor. Upon the receipt of the beacon message, the selected sensor transmits its data packets to the UAV, along with  the state information of $AoI^i_j(t)$ in the control segment of the data packet. After the UAV correctly receives the data, it sends an acknowledgement to the ground sensor.

The following equation highlights the mean field idea of MF-HPPO \cite{yang2018mean}:
\begin{multline} \label{36}
    Q_{i}(s_{\alpha,i},a)=\frac{1}{N_i}\Sigma_{k\in N(i)} Q_i(s_{\alpha,i},a_i,a_k)=\\Q_i(s_{\alpha,i},a_i,o_i).
\end{multline}
Here, $Q_i$ is the $Q$ value of agent $i$, $a$ represents the joint action of all agents. The neighbor agents of agent $i$ are characterized by $N_i$. $o_i$ is an indicator of the mean field.
In essence, in multi-agent systems the Q value of an agent is computed based on the current state and joint action, but when we have a large number of agents computing joint action is impractical, therefore (\ref{36}) allow an agent to compute its Q value just based on the mean field of its neighbors. 

Fig. 2 shows the proposed MF-HPPO with LSTM layer, where each UAV equipped with the MF-HPPO to minimize the average AoI by optimizing the trajectory and data collection schedule.
The use of the LSTM layer, continuous and discrete actors, and the objective function of PPO, are the features of the MF-HPPO in this diagram. As shown, The decision-making component of each agent consists of two actors and a critic, which is preceded by the LSTM layer to draw conclusions based on experience. The actor for continuous action spaces outputs  continuous values for cruise control, such as position and velocity, and the actor for discrete action spaces outputs a categorical value that can be used to select one of the ground sensors. Each agent samples the actions and performs in the environment. The rollout buffer is filled with data generated by these interactions such as, state, mean field, action, cost and policy. As can be seen, we use  Generalized Advantage Estimate (GAE) \cite{schulman2015high} as a sample-efficient method to estimate the advantage function. As depicted, based on the RolloutBuffer, mini-batches are then formed to train the LSTM and the actors and critics so that the agent can continuously improve its policies. The definition of the objective function of PPO is the total of actor losses and critic loss subtracted by entropy, as depicted in the diagram. The actor loss is inputted by the ratio of old policy and current policy and the advantage value. The critic loss is inputted by the critic's output and the return value. The policy is designed to encourage the agent to take advantageous actions, while punishing actions that deviate from the current policy.
\par
Algorithm \ref{al:1} summarizes the MF-HPPO with the LSTM-based
characterization layer. In the initialization step, Input and Output are characterized; the algorithm receives parameters like Clip threshold, discount factor and mini-batch size as input and specify its output as trajectory and scheduling policy of UAV $i$. Next, the actor $\pi_i$  and critic $w_i$ are initialized with random weights for each agent. The number of training episodes is $M$, where the length of each episode is $L$.
Each agent is trained using a predetermined set of iterations throughout the learning phase. Sampling and optimization constitutes the learning phase. In the beginning of learning, the state $s_{\alpha,i}$ and mean field $o_i$ are randomly  initialized for each agent. With the start of the sampling policy, UAV $i$ samples its action based on the policy $\theta_{old}^i$. The sampled action represents sensor selection, velocity and locations, and executed in the environment to obtain the cost, new state and new mean field. Consequently, trajectories (i.e., sequence of states, actions, policy, mean field, and costs) are gathered and stored in the RolloutBuffer. In addition, GAE is applied to calculate the advantage that is used in (\ref{eq:37}). In the optimization step, the policies are optimized. In the optimization step, the policy parameter is updated for each epoch. The PPO objective is computed in each epoch according to the following equation:
\begin{equation} \label{eq:37}
          \begin{split}
             & L^{clip}(\theta^i)=\min\Bigl(\frac{\pi_{\theta^i}(a_i|s_{\alpha,i},o_i)}{\pi_{\theta_{old}^i}(a_i|s_{\alpha,i},o_i)}A_{\pi_{\theta_{old}^i}}(s_{\alpha,i},o_i,a_i),
             \\
             & g(\epsilon,A_{\pi_{\theta_{old}^i}}(s_{\alpha,i},o_i,a_i))\Bigr)
          \end{split}
\end{equation}
where
\begin{equation} \label{eq:100}
    \pi_{\theta^i}(a_i|s_{\alpha,i},o_i)=\pi_{\theta^i}^c(a_i^c|s_{\alpha,i},o_i)\pi_{\theta^i}^d(a_i^d|s_{\alpha,i},o_i).
\end{equation}
\begin{algorithm} [h] 
\caption{MF-HPPO Characterized by LSTM Layer}  \label{al:1}
\textbf{1.Initialize}
\\
\KwIn {Clip threshold $\epsilon$, discount factor $\gamma$, learning rate $\eta$, buffer size $D$, mini-batch size $B$ }
\KwOut {The scheduled ground sensor j and trajectory $\zeta_i$ of UAV i}
Randomly initialize the Actors $\pi_i$ and Critics $w_i$ with networks parameters $\theta^i$
\\
The LSTM layer with $\{W_o, W_c, W_f, W_p\}$ and $\{e_o, e_c, e_f, e_p\}$.
\\
Initialize the sampling policy $\pi_{\theta_{old}^i}$ with
 $\theta_{old}^i \gets \theta^i $.
\\
 $\forall i \in (1,I)$
\\
\textbf{2.Learning}
\\
\For {episode=1 \KwTo $M$}{
      
      Randomly obtain the initial state $s_{\alpha,i}$  
      \\
     \For {$t=1$ \KwTo $L$}{
       \hspace{2cm}\textbf{*The sampling phase*}
       \\
       Sample: Sample action $a_i \sim \pi_{\theta_{old}^i}(a_i|s_{\alpha,i},o_i,\theta^i)$;
       \\
       Execute the action $a_i$ that specifies
       the scheduled ground sensor j  and trajectory $\zeta_i$ of  UAV i.
       \\
       Obtain the cost and new state $s_{\beta,i}$ and new mean field $\o_i(t+1)$.
       \\
       RolloutBuffer: store the trajectory $( s_{\alpha,i},a_i,c,o_i,\pi_{\theta_{old}^i}(a_i|s_{\alpha,i},o_i,\theta^i))$
       \\
       $s_{\alpha,i}=s_{\beta,i}$
        }
          Compute the advantage using GAE
          \\
         \For {$epoch=1$ \KwTo $P$}{
         \begin{flushright}
       \end{flushright}  
         \hspace{2cm}\textbf{*The optimization phase*}
       
         Sample the RolloutBuffer
         \\
         Compute the PPO-Clip objective function using ($\ref{eq:37}$) 
         \\
         Compute the critic loss.
         \\
         Optimize the overall objective function using ($\ref{eq:38}$) 
                                   }
    Synchronize the sampling policy
    $\pi_{\theta_{old}^i}  \gets \pi_{\theta^i}$ 
    \\
    Drop the stored data in RolloutBuffer.
   }
\end{algorithm}
Here $a_i^c$ and $a_i^d$ correspond to actions in continuous and discrete spaces. In (\ref{eq:100}), to obtain the hybrid policy $\pi_{\theta^i}(a_i|s_{\alpha,i},o_i)$, we multiply the policies for continuous and discrete actions \cite{neunert2020continuous}. Meanwhile, we assume that wireless radio of the UAV can cover the whole field.

Continuous policy $\pi_{\theta^i}^c$ is modeled using multivariate normal distribution and discrete policy $\pi_{\theta^i}^d$ is modeled using categorical distribution. 
In the next step, the overall objective function is optimized according to the following equation:

\begin{equation} \label{eq:38}
          L^{total}(\theta^i)=L^{clip}(\theta^i)-K_1 L^{VF}(\theta^i)+K_2*H.
\end{equation}
\begin{figure*}[ht] 
        \centering
        \captionsetup{justification=centering}
    	\includegraphics[ width=5in]{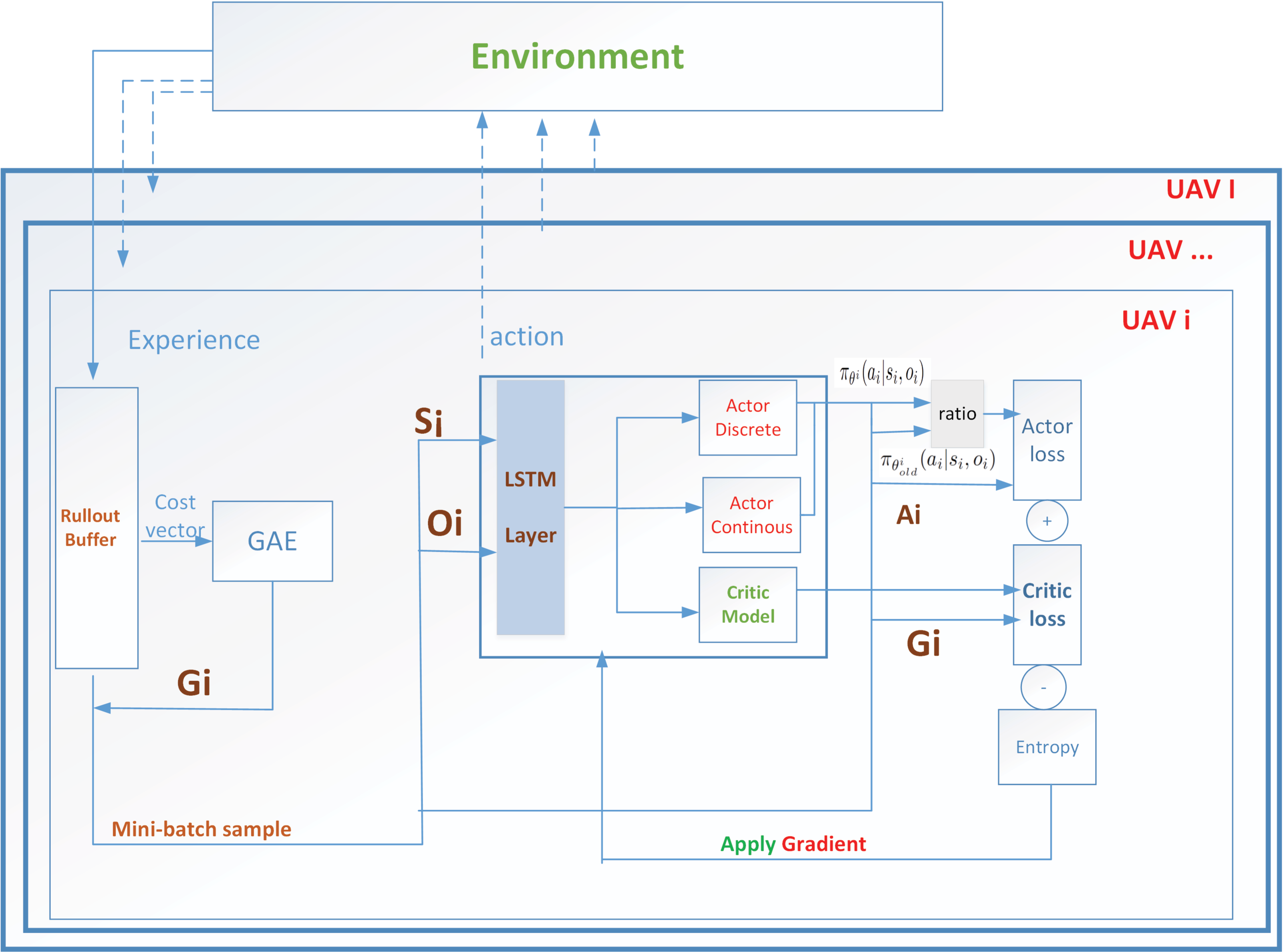}
    	\caption{An overview of MF-HPPO, where each UAV is equipped with the LSTM layer to optimize discrete and continuous actions using hybrid policy. 
}
    	\label{fig:MF-HPPO}
\end{figure*}

\noindent Here, $L^{VF}(\theta^i)$ is the critic loss and $H$ acts as a regularizer encourages the agent to execute actions more unpredictably for exploration and guard against the policy being overly deterministic. The entropy for continuous and discrete actions is computed based on the actions' distribution. We obtain the entropy by  multiplication of the entropy of continuous and discrete action spaces to enable enforcing consistent regularization to both continuous and  discrete action spaces. $K_1$ balances the importance of the critic loss and the actor loss, and $K_2$ coefficient controls the amount of entropy in the policy. 

Finally, the sampling policy $\pi_{\theta_{old}^i}$ is updated with the policy $\pi_{\theta^i}$, and the
stored data are dropped. The next iteration then begins.The proposed MF-HPPO model is driven by DRL and can improve data aggregation by learning and refining the actions of cruise control and communication schedules on the fly. By this means, the MF-HPPO model can account for generic collision or obstacle avoidance via the offline training and can adapt to the specific real-world application scenario via online refinement. The UAVs can also be equipped with event cameras or utilize vision-based techniques to avoid collisions and adjust the flight behavior \cite{9981803}, \cite{9108245}. 
\subsection{LSTM Layer} \label{sub:54}
We further develop an LSTM layer in the proposed MF-HPPO, which captures long-term dependencies of time-varying network state $s_\alpha$. Cell memory and the gating mechanism are main components of LSTM. Cell memory is responsible to store the summary of the past input data and the gating mechanism regulates the  information flow between the input, output, and cell memory. The network states are fed into LSTM one by one (one at each step). The last hidden state $\kappa_i^{hidd}$  is returned as the output of the state characterization layer.
Each agent uses an LSTM layer to predict their
respective hidden states. The hidden states $\kappa_i^{hidd}$
 are calculated by the following
composite function:
\begin{equation}
    \kappa_i^{hidd}=out_itanh(C_i),
\end{equation}
\begin{equation}
    out_i=\sigma(W_0 \cdot [C_i,\kappa_{i-1}^{hidd},A_i]+e_i ),
\end{equation}
\begin{equation}
    C_i=F_iC_{i-1}+p_itanh(W_c.[\kappa_{i-1}^{hidd},A_i]+e_c),
\end{equation}
\begin{equation}
    F_i=\sigma(W_f \cdot [\kappa_{i-1}^{hidd},C_{i-1},A_i]+e_f),
\end{equation}
\begin{equation}
     p_i=\sigma(W_f \cdot [\kappa_{i-1}^{hidd},C_{i-1},A_i]+e_p),
\end{equation}
where the output gate, cell activation vectors, forget gate, and input gate of the LSTM layer are denoted by $out_i$, $C_i$, $F_i$, and $p_i$, respectively. $\sigma$ and tanh correspond to logistic sigmoid
function and the hyperbolic tangent function, respectively.
${W_0,W_c,W_f ,Wp}$ are the weight matrix, and
${e_0, e_c, e_f , e_p}$ are the bias matrix\cite{9507550}, \cite{9779339}.

\subsection{Complexity and Convergence of MF-HPPO}
The overall complexity of MF-HPPO is calculated as follows,
$O(I \cdot ML \cdot(\Sigma_{g=1}^G n_{g-1}.n_g))$
where $n_g$ is the number of neural units in the g-th hidden layer. In this work, the PPO architecture is built with the same $n_g$ in all hidden layers. Therefore, the PPO complexity can be reduced to $O(I \cdot ML \cdot(g-1) \cdot n_g^2)=O(I \cdot ML \cdot n_g^2)$. The convergence analysis is proved by simulation results (see Fig. \ref{fig:conv}).

\section{NUMERICAL RESULTS AND DISCUSSIONS} \label{sec:6}
\subsection{Implementation of MF-HPPO}
MF-HPPO is implemented in Python 3.8 using Pytorch (the Python deep learning library). A Predator Workstation running 64-bit Ubuntu 20.04 LTS, with Intel Core i7-11370 H CPU @ 3.30 GHz 8 and 16 GB memory is used for the Pytorch setup. Table  \ref{table:2} clearly outlines the
different considered simulation parameters.
MF-HPPO algorithm is trained over 3000 episodes with 40 steps each. The discount factor and learning rate are set
to 0.99 and 3e-4, respectively. Each agent comprises the input layer, LSTM layer, the critic and actors with fully-connected hidden layers of size 256 and output layer. Each neuron uses Rectified Linear Unit (ReLU) as an activation function. In addition, Hyperbolic tangent (tanh) and softmax are used as activation functions in the output layer of the continuous actor-network and discrete actor network. The input of each critic network is represented as a concatenation of states and mean field, and its output is a scalar that assesses the states according to the global policy. The total log probability of the hybrid policy is the sum of the log probabilities of the continuous and discrete action spaces. This log probability would be used as part of the calculation of the objective function in MF-HPPO, along with the estimated cost and the entropy regularization term.
\subsection{Baseline Description}
The MF-HPPO characterized with LSTM layer is compared by single-agent PPO, random scheduling and trajectory design (RSTD), multi-agent DQN (MADQN) and MF-HPPO without LSTM Layer.
\begin{figure*}[ht!]
    \centering
  
    \begin{subfigure}[b]{0.45\textwidth}
        \centering
        \includegraphics[ width=\textwidth]{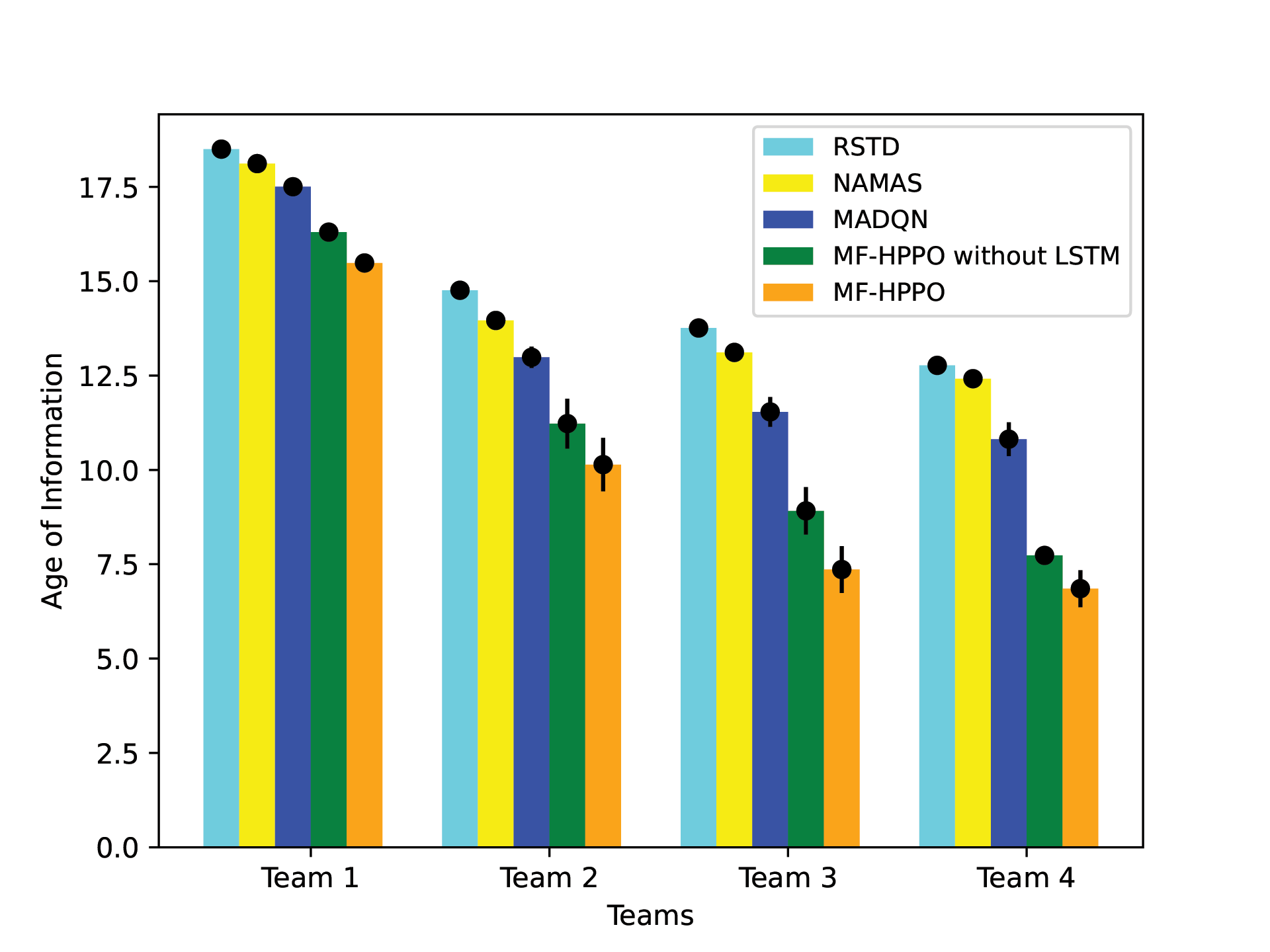}
        \caption{Evaluation of MF-HPPO's performance with a variable number of UAVs in comparison to RSTD, MADQN and MF-HPPO without LSTM}
        \label{fig:multiuavn}
    \end{subfigure}
    \hfill
    \begin{subfigure}[b]{0.45\textwidth}
        \centering
        \includegraphics[ width=\textwidth]{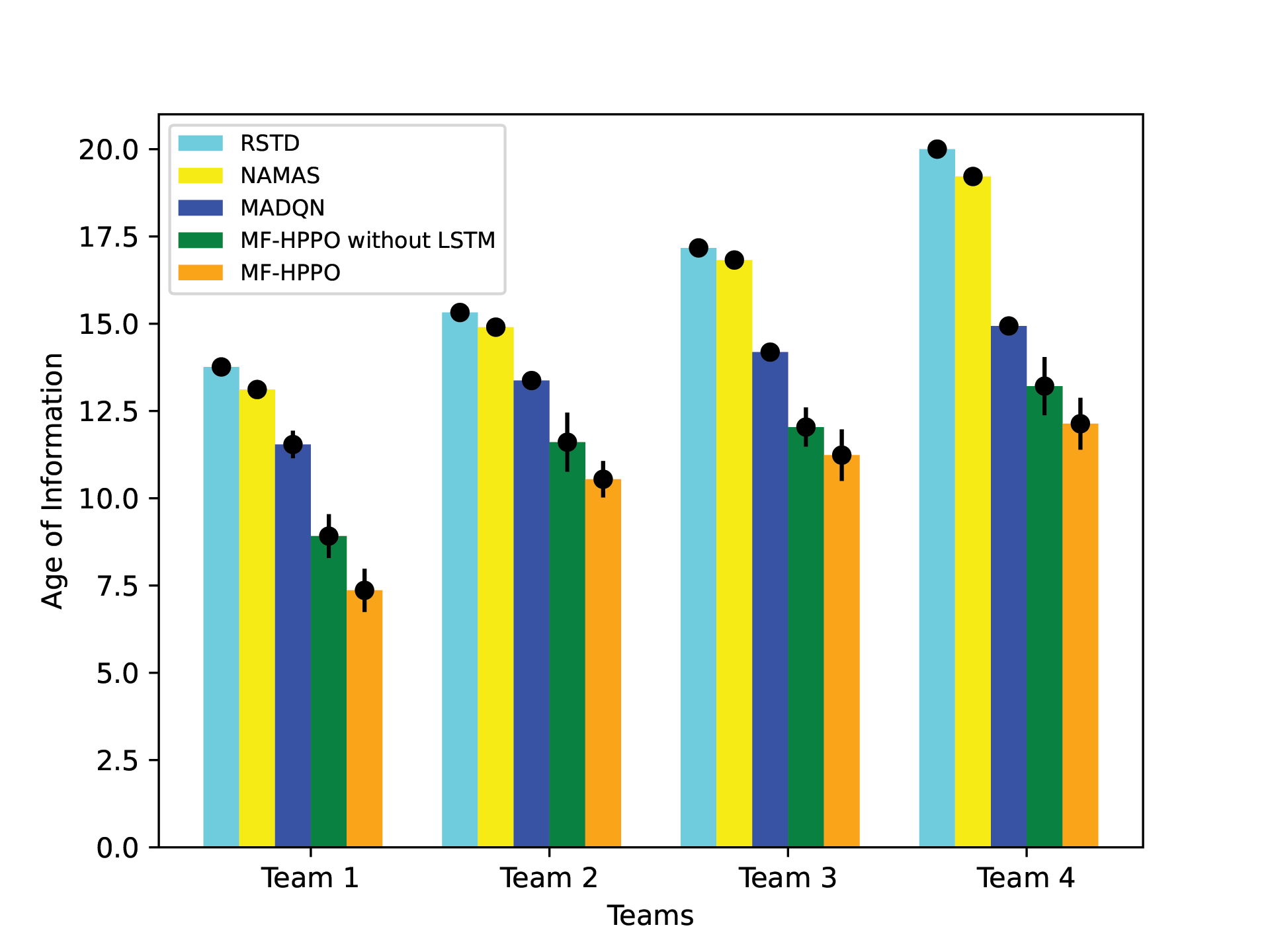}
        \caption{Evaluation of MF-HPPO's performance with a variable number of ground sensors in comparison to RSTD, MADQN and MF-HPPO without LSTM}
        \label{fig:ground sensorsn}
    \end{subfigure}

      \caption{Performance evaluation of MF-HPPO by changing the number of UAVs and ground sensors }
    \label{fig:uavsensor}
\end{figure*}
\begin{figure} [ht!]
        \includegraphics[width=3.5in, height =2.5in]{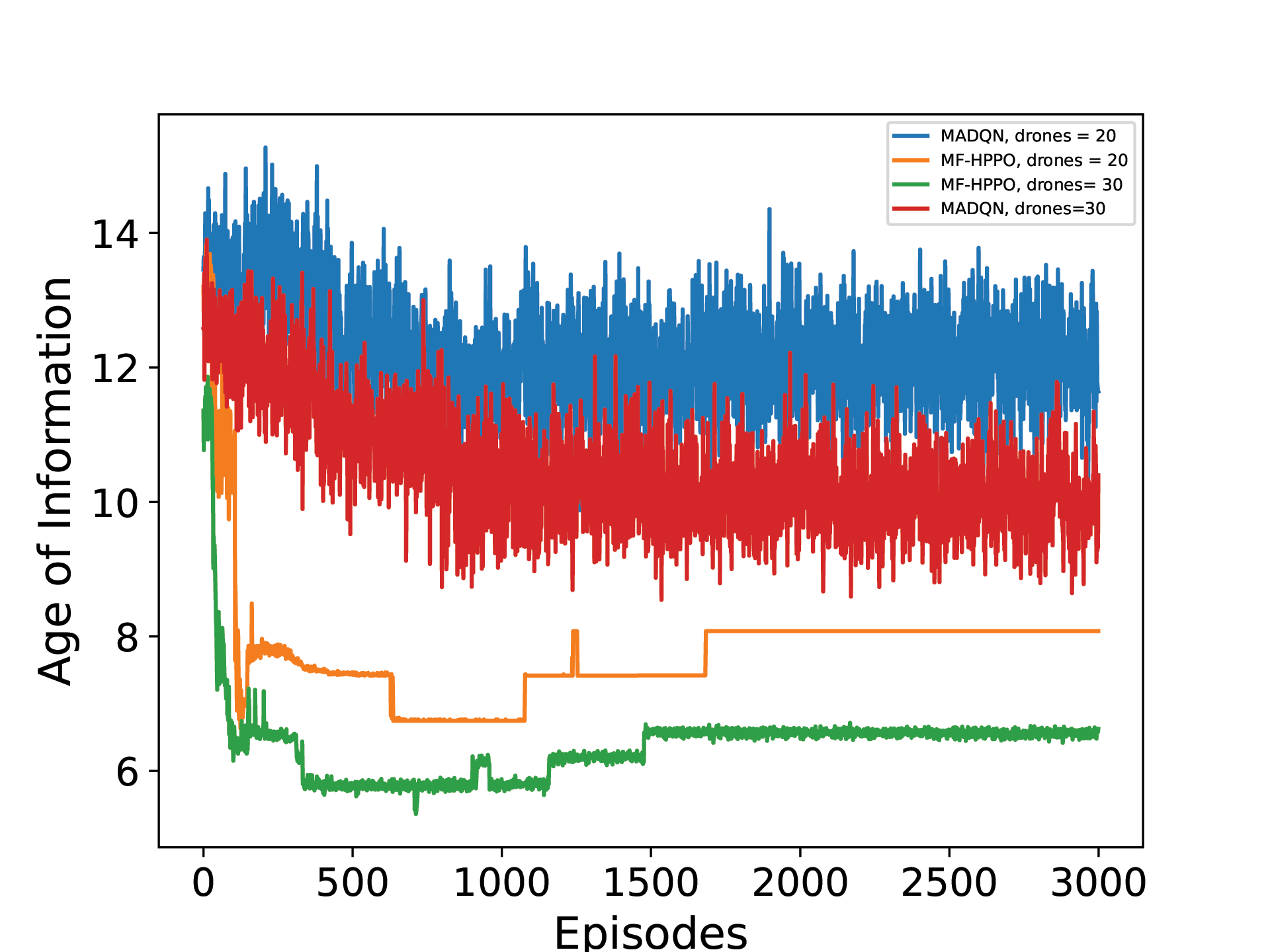}
    	\caption{The network cost for each episode of MF-HPPO with I=30 and benchmarks}
    	\label{fig:conv}
\end{figure}
A brief introduction of the four benchmarks is given below
\begin{enumerate}
    \item NAMAS \cite{9013924}, in this algorithm, each agent schedules the neighbors with maximum AoI to minimize the average AoI.
    \item RSTD, in this algorithm transmission scheduling and trajectory design, are randomly designed. 
    \item MADQN, in this algorithm, each agent running DQN cooperates to reduce average AoI following circular trajectories.
    \item MF-HPPO without LSTM Layer, the structure of this algorithm is the same as MF-HPPO but without LSTM layer.
\end{enumerate}

\begin{table}[t]
\centering
\small
\caption*{ Table II: PyTorch Configuration}
\begin{tabular}{|c|c|} 
 \hline
  \textbf{Parameters} &  \textbf{Values} \\
 \hline
 Number of ground sensors & 100\\ 
 \hline
 Number of UAVs & 30 \\
 \hline
 Geographical area size [m] & 1,000*1,000 \\
 \hline
 Altitude of the UAVs & 120 m \\
 \hline
 Critic Network Learning Rate & 3e-4 \\
 \hline
 Actor-Network Learning Rate & 3e-4 \\
 \hline
 Number of Hidden Layers for Networks & 2 \\
 \hline
 Number of Neurons & 256  \\
 \hline
 Loss Coefficients for $K_1$ and $K_2$ & 0.2 and 3 \\
 \hline
 Optimizer Technique & Adam \\
  \hline
  Clip Fraction & 0.2 \\
  \hline
 Rollout Buffer size & 40 \\
 \hline
 Batch size & 40 \\
 \hline
 Mini Batch Size & 4 \\
 \hline
 PPO Epochs & 8 \\
 \hline
  Number of episodes & 3,000 \\
 \hline
 Discount Factor & 0.99 \\
 \hline
 Minimum speed   & 0 m/s \\
 \hline
 Maximum speed  & 15 m/s \\
 \hline
\end{tabular}
\label{table:2}
\end{table}
\subsection{Performance analysis of MF-HPPO }

Fig. \ref{fig:uavsensor} depicts the performance evaluation of MF-HPPO in comparison to the baselines by changing the number of UAVs and ground sensors. Fig. \ref{fig:multiuavn} illustrates the influence of varying the quantity of UAVs on the AoI. It is observed that an increase of UAVs leads to a reduction in AoI, attributable to enhanced time efficiency and the capacity for quicker operation of more ground sensors. Specifically, augmenting the UAV count from 1 to 30 results in a 61\% reduction in the average AoI for the MF-HPPO algorithm, in contrast to a 37\% reduction observed for MADQN. This disparity is attributed to the fact that MF-HPPO executes optimization within a mixed action space, demonstrating greater training stability compared to MADQN, which utilizes circular trajectories. Furthermore, it is noteworthy that MF-HPPO significantly surpasses the performance of both random assignment and NAMAS strategies.
Fig. \ref{fig:ground sensorsn}  evaluates the average AoI given 20 UAVs and groups of 100, 200, 300, and 400 ground sensors. The MADQN, NAMAS, and the RSTD are used as baselines. Overall, increasing the number of ground sensors results in a uniform increase in the average AoI, since more sensor data should be collected. In particular, when the number of ground sensors is 400, the proposed MF-HPPO outperforms the RSTD by 38\%, NAMAS by 33\%, and the MADQN by 17\%.

Fig. \ref{fig:conv} captures the convergence trend of the MF-HPPO algorithm, which was assessed by deploying 20 UAVs to service 100 ground sensors. In this context, the MF-HPPO model with I=30 settings demonstrates a significantly lower Age of Information (AoI) when compared to the MADQN algorithm with I=20 and I=30, exhibiting improvements of 38\% and 43\%, respectively. This enhanced performance is attributed to the optimized trajectories and scheduling for data collection by multiple UAVs, resulting in superior time efficiency. Additionally, the integration of an LSTM layer in the MF-HPPO framework contributes to both an acceleration and stabilization of convergence. Notably, the peak AoI for the proposed MF-HPPO model decreases dramatically from 14 seconds to 6 seconds within the initial 1,000 episodes. Subsequently, between episodes 1,500 and 3,000, the AoI stabilizes at around 7 seconds, with only minor fluctuations observed
\begin{figure*}[ht!]
    \centering
  
    \begin{subfigure}[b]{0.32\textwidth}
        \centering
        \includegraphics[ width=\linewidth]{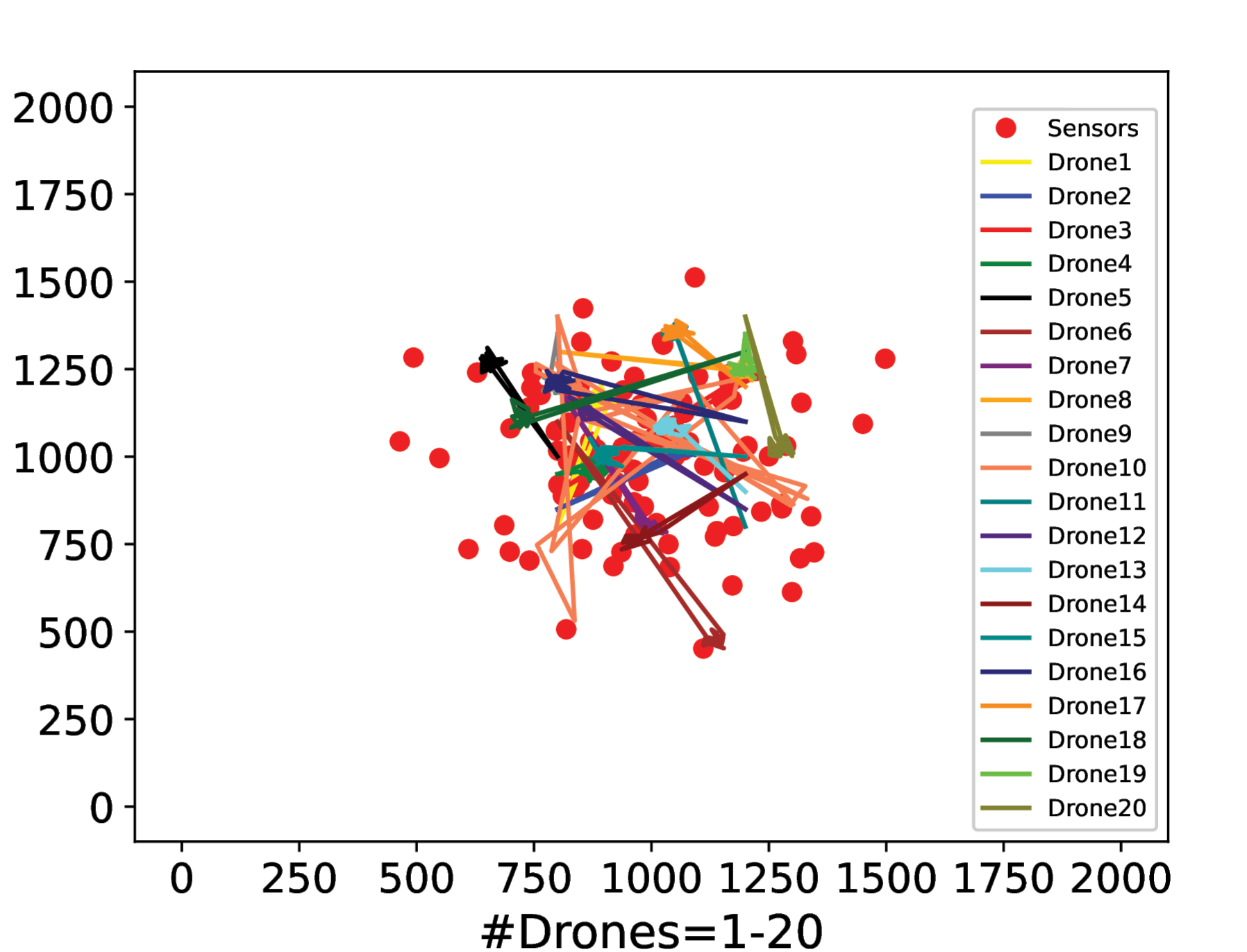}
        \caption{Normal Distribution}
        \label{fig:velocity20}
    \end{subfigure}
    \hfill
    \begin{subfigure}[b]{0.32\textwidth}
        \centering
        \includegraphics[ width=\linewidth]{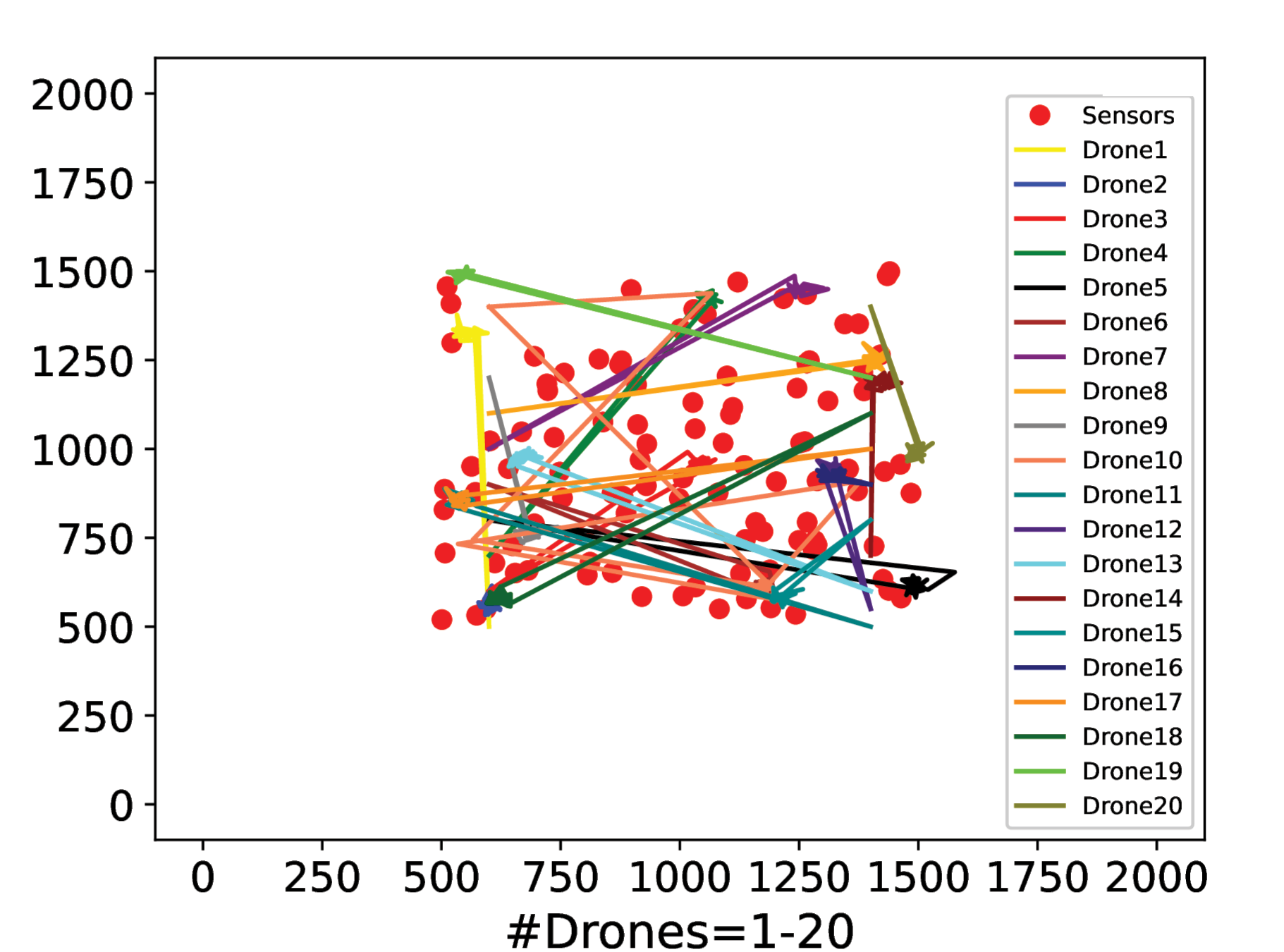}
        \caption{Square Distribution}
        \label{fig:trajectory20}
    \end{subfigure}
    \hfill
    \begin{subfigure}[b]{0.32\textwidth}
        \centering
       
        \includegraphics[width=\linewidth]{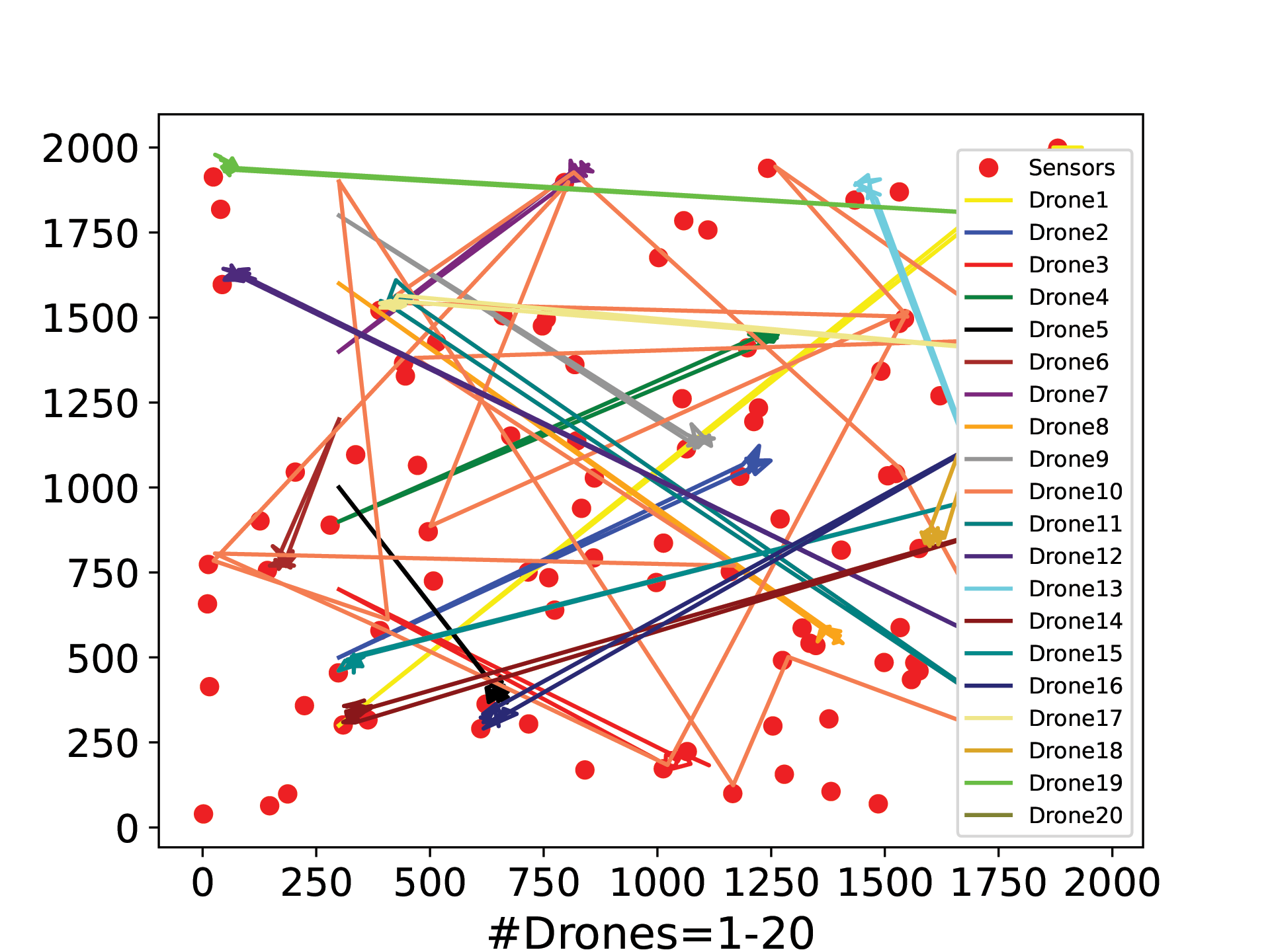}
        \caption{Uniform Distribution}
        \label{fig:trajectory20}
    \end{subfigure}
    \caption{MF-HPPO trajectory distributions for various UAV counts and ground sensor distributions. }
    \label{fig:four figures}
\end{figure*}

\begin{figure} [ht!]
        \includegraphics[width=3.5in, height =2.5in]{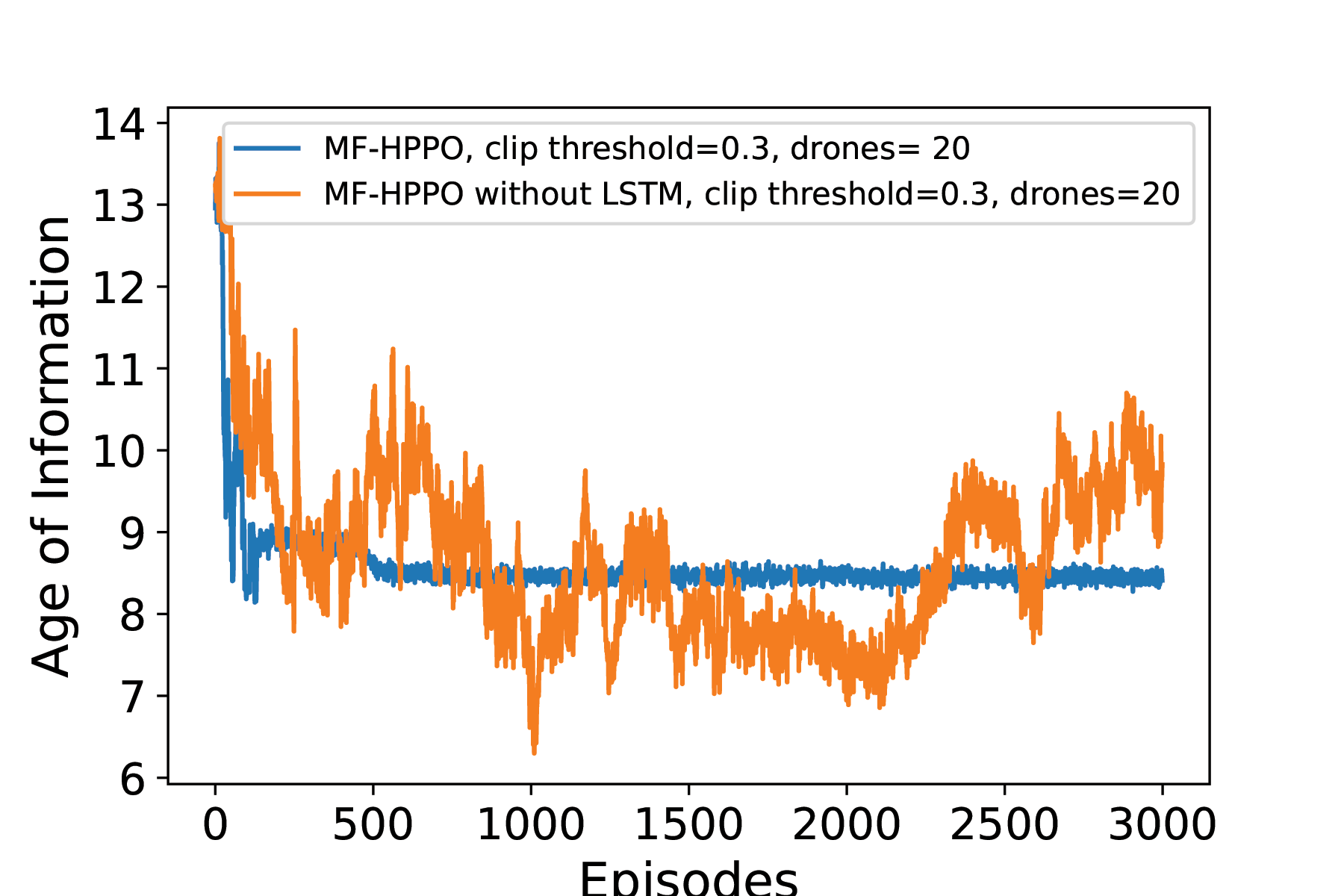}
    	\caption{Performance evaluation of MF-HPPO by changing clip threshold}
    	\label{fig:clip}
\end{figure}

MF-HPPO-generated trajectories for 20 UAVs are shown in Fig. \ref{fig:four figures}, where the ground sensor distribution patterns are uniform, square, or normal ones.
When designing trajectories for AoI minimization, the UAVs' trajectories are impacted by the distribution of the ground sensors. The UAV needs to approach to the location of each scheduled sensor to collect the data and update its AoI.
Fig. 5(a), refer to the normal distribution and shows  trajectories for 20 UAVs, focusing on the center area of the ground sensors and less on the corners. The normal distribution of the ground sensors can affect the UAVs' trajectories by determining which ground sensors are prioritized for data collection. For example, as can be seen, most ground sensors are centered and their data may become stale, in this case, the UAVs'trajectories are designed to visit these ground sensors more frequently to minimize the average AoI. Figs. 5(b) is related to the square distribution. As can be seen, the ground sensors are less centered. This cause diverse set of ground sensors in wider range to be covered in comparison to normal distribution. Fig. 5(c) refer to the uniform distribution. As can be seen, the UAVs design wide-area trajectories due to the wider distribution of ground sensors covering the entire area and the AoI requirements of the scattered ground sensors.

Fig. \ref{fig:clip} demonstrates the convergence figures for two variants of MF-HPPO by changing the clip threshold.
PPO uses the clip threshold, commonly referred to as epsilon, to regulate the amount of policy updating. A larger clip threshold allows for more aggressive updating, while a smaller clip threshold restricts updating more severely, resulting in less policy change. The blue curve shows the MF-HPPO with LSTM layer and a clip threshold of 0.3 outperforming the MF-HPPO without LSTM layer clip threshold 0.3. The latter shows a deviating behavior due to the influence of the clip threshold, while the blue curve shows an absolutely stable trend despite the same value of the clip threshold thanks to the LSTM layer. Overall, adding the LSTM layer to MF-HPPO can stabilize the training and prevent divergence of the strategies.

\section{Conclusion} \label{conc}
In this paper, we propose a mean field flight resource allocation to model velocity control for a swarm of UAVs, in which each UAV minimizes the average AoI by considering the collective behavior of others. Due to the high computational complexity of MFG, we  leverage AI and propose MF-HPPO characterized with an LSTM layer to optimize the UAV trajectories and data collection scheduling in mixed action space. Simulation results based on PyTorch deep learning library show that the proposed MF-HPPO for UAVs-assisted sensor networks reduces average AoI by up to 57\% and 45\%, as compared to existing non-learning random algorithm and MADQN method (which performs the action of trajectory planning in the discrete space), respectively. This confirms the AI-enhanced mean field resource allocation is a practical solution for minimizing AoI in UAV swarms.
\appendix
\section*{Proof of FPK Equation (\ref{eq:13}) for Cruise Control}
We derive the mean field via an arbitrary test function $g(\zeta)$, which is a twice continuously differentiable compactly supported function of the state space. The integral of $m(\zeta)g(\zeta)d\zeta$ can be considered as the continuum limit of the sum $g(\zeta(t))$, where $\zeta(t)$ is the UAV's state at time $t$. It is known that,

\begin{equation}  \label{eq:41}
    \int m(\zeta(t)) g(\zeta) d\zeta=\frac{1}{N}\Sigma_{i=1}^N g(\zeta(t)).
\end{equation}
At time $t$, the first-order differential function with regard to time $t$ is derived to check how this integral varies in time. By utilizing the chain rule, we can derive the heuristic formula as	
\begin{multline} \label{eq:42}
    \int \partial_t m(\zeta(t)) g(\zeta) d\zeta=\\ \frac{1}{N}\Sigma_{i=1}^N \partial_t \zeta(t)\nabla g(\zeta(t))+\partial_t^2 \zeta(t) \nabla^2 g(\zeta(t)).
\end{multline}
Taking the limit of the right side of the above equation when $N$ tends
to infinity, we get 
\begin{multline}  \label{eq:43}
    \int [ \partial_tm(\zeta(t))+\nabla_{\zeta}m(\zeta(t))\cdot \frac{\partial \zeta}{\partial t}- \\ \frac{\eta^2}{2}\nabla_{\zeta}^2m(\zeta(t))] g(\zeta(t))d\zeta=0,
\end{multline}
for any test function $g$ through integration by parts. Then the above equation leads to the following equation:
\begin{equation} \label{eq:44}
    \partial_tm(\zeta(t))+\nabla_{\zeta}m(\zeta(t))\cdot v(t)- \frac{\sigma^2}{2}\nabla_{\zeta}^2m(\zeta(t))=0. 
\end{equation}
which correspond to FPK equation defined in (\ref{eq:13}).
\bibliographystyle{IEEEtran}
\bibliography{references}
\end{document}